\documentclass[10pt,a4,twocolumn]{article}

\usepackage{amssymb}
\usepackage[cmex10]{amsmath}
\usepackage{hyperref}
\usepackage[utf8]{inputenc}
\usepackage[letterpaper,top=1.5in,bottom=1.5in,left=1in,right=1in]{geometry}
\usepackage{graphicx}
\usepackage{subfigure}

\newcommand{\reffig}[1]{Fig.~\ref{#1}}

\newcommand{\reftbl}[1]{Table~\ref{#1}}

\newcounter{cntrObservation}

\newcommand{\hCountDiff}{\stepcounter{cntrObservation}\textbf{Observation~\arabic{cntrObservation}.~}}

\usepackage{enumitem}

\begin{document}

\title{%On the Gender Difference
Asymmetries of Men and Women in Selecting Partner}
\author{Haluk O. Bingol and Omer Basar\\
%\thanks{Thanks to the editors of this wonderful journal!}\\
\small Department of Computer Engineering\\[-0.8ex]
\small Bogazici University, Istanbul\\
\small \texttt{bingol@boun.edu.tr}\\
%\and
%Forgotten Second Author\\
%\small School of Hard Knocks\\[-0.8ex]
%\small University of Western Nowhere\\[-0.8ex]
%\small Nowhere Uvherdov\\
%\small \texttt{no1remembers@me.woe.edu}
}

%\author{Haluk O. Bingol}
%\affiliation{
%	Department of Computer Engineering\\
%	Bogazici University, Istanbul \\
%	\today 
%}
%\date{\today} 

%\pacs{89.75.Hc, 89.65.Ef, 89.75.Fb}

\maketitle

\begin{abstract}
This paper investigates human dynamics in a large online dating site with 
3,000 new users daily who stay in the system for 3 months on the average. 
The daily activity is also quite large such as 
500,000 massage transactions, 
5,000 photo uploads, and 
20,000 votes. 

The data investigated has $276,210$ male and $483,963$ female users.
Based on the activity that they made, 
there are clear distinctions between men and women in their pattern of behavior.
Men prefer lower, women prefer higher qualifications in their partner.

\end{abstract}

% ------

\section{Motivation}

Until very recently we all live in a physical world only. 
With the rapid penetration of Internet to our daily lives starting in mid 1990's,
we developed a new concept of virtual world.
Now we have virtual friends that we exchange emails or tweets.
We used to live in our social network, 
which was within our physical proximity. 
Now we talk about friends that we have not seen the face but 
know all of her ideas in almost all issues since 
we read his communications in the email list, or in her blog, or by means of her tweets.

Thanks to mobile Internet and new mobile devices 
such as smart phones and tablets
we now have the ability to be connected to this virtual world 
7-day-24-hour basis.
%Especially kids are well adapted to this new world.
%With massive player games, they prefer to play virtual rather than physical.

This is a new phenomenon that has never happened before.
The impact of information and communication technologies (ICT) changed 
%the way we live our lives~\cite{Salah2011HBU,Lepri2011AMI}.
the way we live our lives~\cite{Salah2011HBU}.
The usual patterns of conducting our lives are in transition to adopt to the new world of ICT.
On the other hand, it provides invaluable datasets that have never been possible 
before~\cite{
	Dodds2003Science, 
	%Barabasi2005Nature, 
	Leskovec2006LNCS,
	Onnela2007PNAS,
	Barraket2008JSociology,
	Rybski2009PNAS,
	Centola2010Science,
	Rocha2010PNAS,
	Szell2010PNAS,
	%Basar2010,
	Centola2011Science}.
Dating is not that can be kept away from this trend of 
%virtualization~\cite{Barraket2008JSociology,Basar2010}.
virtualization~\cite{Barraket2008JSociology}.

Men and women are different in many ways~\cite{
	Silverman1992,
	Liljeros2001Nature,
	CelaConde2009PNAS,
	Hyde2009PNAS}.
First of all they are different genetically due to $X$ and $Y$ chromosomes. 
One theory to explain genetic difference is the hunter-gatherer theory of Silverman which 
proposes a division of labor between genders, that is, 
men are hunters and women are foragers~\cite{Silverman1992}.
It has been shown that the brain of different genders works differently.
For example, 
an electrophysiological study shows that 
men and women use different strategies, 
hence, different parts of their brain 
while judging beauty~\cite{CelaConde2009PNAS}.  
Men use mainly the right hemisphere  for coordinate-based strategies.
On the other hand women use both hemispheres for categorical strategies.
They behave differently in many occasions.
For example, men tent to have more sexual partners then women~\cite{Liljeros2001Nature}.

There are differences in gender 
but some of these differences may be due to social reasons 
such as historical or cultural.
In late 1800, women are thought to be inferior to men in analytic abilities.
As women get more share of everyday life, 
this view is changing.
There is a clear trend of improvement of girls in test scores, such as SAT or PISA, as they take more science courses in the education system~\cite{Hyde2009PNAS}.
Once male dominated STEM fields, 
that is, the fields of science, technology, engineering, and mathematics, 
is becoming more even in genders.

Data for gender difference studies come from a number of resources. 
A survey based data collection is a method that is used frequently. 
As in any survey, the answers may be biased. 
Another data collection is large-scale activities such as SAT tests. 
Recently the trace that we all leave on Internet becomes a source of data. 
In this paper, we investigate the data of a relatively large online dating system with focusing on the difference between men and women.

\section{Online Dating Systems}

A typical online dating system enables its user to 
find partner that best matches one's desires.
Each user defines his \emph{user profile} in which 
he describes himself in terms of answering a set of questions 
such as gender, age, education, and income level.
The questions are design in such a way that 
the answers are usually in the form of multiple choice 
rather than free format entries.
User is asked to select one choice out of many.

A user searching for a partner defines 
what properties that he wants in his partner.
This profile is called the \emph{desired profile}.
The same set of questions for profiling is used for this. 
The only difference is that 
one may select many out of multiple choices.
For example, in his user profile he sets his education level as ``university''
while he can select 
``high school'', ``masters'' as well as ``university'' in the desired profile.
Then all the users with profiles that match this are potential candidates.

\section{The Dataset}

The data that we investigate is based on an online dating system 
that is well established and one of the largest in Turkey.%~\cite{Basar2010}.

There are 4,500,000 users registered.
One should be careful about this big number.
It is known that a user uses the system for three months on the average, 
than leaves it.
Later, he may come back with his old user account. 
It is also possible that he creates a brand new account.
There is no way for us to identify this new account to the old one.
Therefore, this big number should not be understood as that many individual users.
On the other hand, 
the system is used quite extensively.
More than 3,000 new users registered daily.
More than 50,000 users log on to the system daily.

The data investigated is huge.
Considering the daily activity given in \reftbl{tbl:DailyActivity}
and 
this activity is accumulated through the years,
the data size is in the order of couple of 100 GB.

The possible activities that users can make is also listed in \reftbl{tbl:DailyActivity}.
User can \emph{wink} to a user to check if there is a potential interest.
User can sent a \emph{message} that is one-step higher involvement than a wink.
It is possible that a \emph{virtual gift}, images of flowers, can be sent.
The difference of gifts from other actions is that 
the gifts are individually purchased 
while other actions are included in the membership fee.
Therefore gift some how established a special value in the user community.
A user can put another user in his \emph{favorites} list.
He can also \emph{vote} to mean that he would go for her.

\begin{table}
\begin{center}
	\caption{
		Daily activity numbers 
	}
	\begin{tabular}{|l|r|}
	    \hline
	    Activity &Number\\
	    \hline  
		Messages sent &500,000\\
		Gifts sent &1,000\\
		Winks &10,000\\
		Favorites &10,000\\
		Votes &20,000\\
		Photos uploaded &5,000\\
		New users &3,000\\
		Logins &50,000\\
	    \hline
	\end{tabular}
	\label{tbl:DailyActivity}
\end{center}
\end{table}

We investigate the data of 
$\mathcal{M} = 276,210$ male and 
$\mathcal{F} = 483,963$ female users.
It may be surprising to see that there are more female in the data set, 
which calls for explanation.
Due to the business model of the site, 
the system is free of charge for female users. 
On the other hand, 
there are two types of male users.
The \emph{free-of-charge male users} have restricted access to the facilities of the site.
In order to get full access to the facilities of the site, 
male has to be a \emph{paid member}.
Our investigations require full access to facilities for both male and female users. 
Hence, we investigate all female users and only male users with paid membership.
Since the free male users are not included, 
the number of male users is much less than that of female users. 

As the investigation goes, 
some of the users do not satisfy the question at hand.
Then the size of the sample space, that we get the statistics, reduces.
For example, 
when we investigate the behavior of user that already sent a virtual gift twice, 
the size of the sample would be less that
that of the sample who both sent it twice and 
having an educational level of high school or higher.
Since the number of male and female users is well above a few thousands 
even for the most restricted queries that we use, 
statistical values expected to be well founded.
Since the number of female and male considered for a particular question 
changes by the question, 
we provide the number of male and female participant for that particular case by 
$\mathcal{M}$ and $\mathcal{F}$ values, respectively.

Clearly, privacy is the most important issue for such investigation.
No individual personal information is used in this study.
Data is randomized with MD5 algorithm for privacy concerns.
No data left the company. 
All the data processing is done at the site of the company.
Only statistical aggregated data is investigated in this study.

\section{Observations}

\subsection{Profile}

The mandatory fields of profile are 
birthday, 
marital status, 
town that you live in, 
your gender, and 
gender that you are
looking for.
Among the mandatory fields, 
the age has special importance
since users under 18 are not allowed to use such systems by law.

Other properties of the profile are optional.
User may or may not provide answers to them.
A rich set of optional questions help user to define themselves.
Among those are 
height, 
weight, 
eye color, 
hair color, 
body type, 
education level, 
occupation, 
position at work, 
salary,
foreign languages, 
with whom he is living,
parenthood status, 
child desire,
smoking,  
alcohol status,
as well as,
the relation type 
that is sought for.

One needs to be careful that the profile is based on user's claims, 
that is, 
it may or may not be the real value of the user. 
On the other hand, cheating on the properties too far would be not a good strategy 
since when the time comes to meet face to face unfaithful declaration, 
such as declared as slim while being obese, 
would be obstacle to further the relationship.
So we assume that user are closed to what they claim to be.

\begin{figure}%[ht]
\begin{center}
	\includegraphics[scale=0.5]{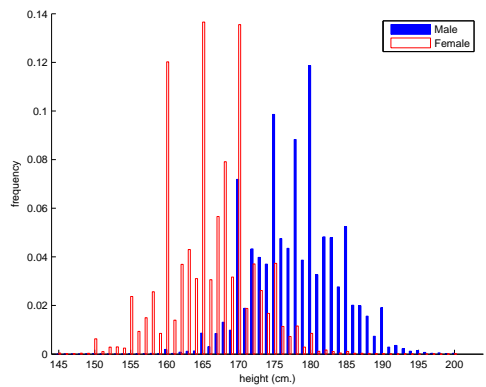}
	\caption{
		Users round up their height to the numbers multiple of 5.
		($\mathcal{M}$ = 276,210, $\mathcal{F}$= 483,963.)
	}
	\label{fig:profileHeight}
\end{center}
\end{figure}

\hCountDiff
Although the system enables to select the exact height such as 178 cm,
it is observed that 
users round up their height to the numbers multiple of 5, such as 170, 175, and 180.
In \reffig{fig:profileHeight} we see jumps at the multiples of 5.
Clearly there is a manipulation of information by the user.
One may interpret this, 
as user prefers to use next level that 
he thinks better for him for the purpose.
This is the case for both genders.

\hCountDiff
The first behavioral difference of the two genders
that we observed is the percentage of the population that answers the profile questions.
Although they are not required, 
majority of the users prefer to declare these fields. 
There is a clear difference in gender as seen in \reftbl{tbl:PercentagesOfAnswers}.
Male does not hesitate to share information
where as female is more concerned.
For age, education and body type they behave almost the same
but for height the difference starts.
For salary, the percentage of male is almost twice of that of female.

\begin{table}
\begin{center}
	\caption{
		Percentages of answers by gender 
		($\mathcal{M} = 276,210$, $\mathcal{F} = 483,963$)
	}
	\begin{tabular}{|l|r|r|}
	    \hline
	    Property &Male &Female\\
	    \hline  
	    Age &100 &100\\ 
	    Education &99 &97\\ 
	    Body type &99 &97\\ 
	    Height &98 &88\\ 
	    Salary &65 &35\\ 
	    \hline
	\end{tabular}
	\label{tbl:PercentagesOfAnswers}
\end{center}
\end{table}

\subsection{Distribution of Property Difference}

It is generally believed that 
people interact with people who have similar 
%characteristics~\cite{McPherson2001,Centola2007c,Centola2011Science}.
characteristics~\cite{McPherson2001,Centola2011Science}.
In the context of this investigation, 
that means that 
someone, who claims to be $170$ cm tall, 
is expected to become friend  with the similar height.

\begin{figure}%[ht]
\begin{center}
	\subfigure[Height]{
		\includegraphics[scale=0.34]{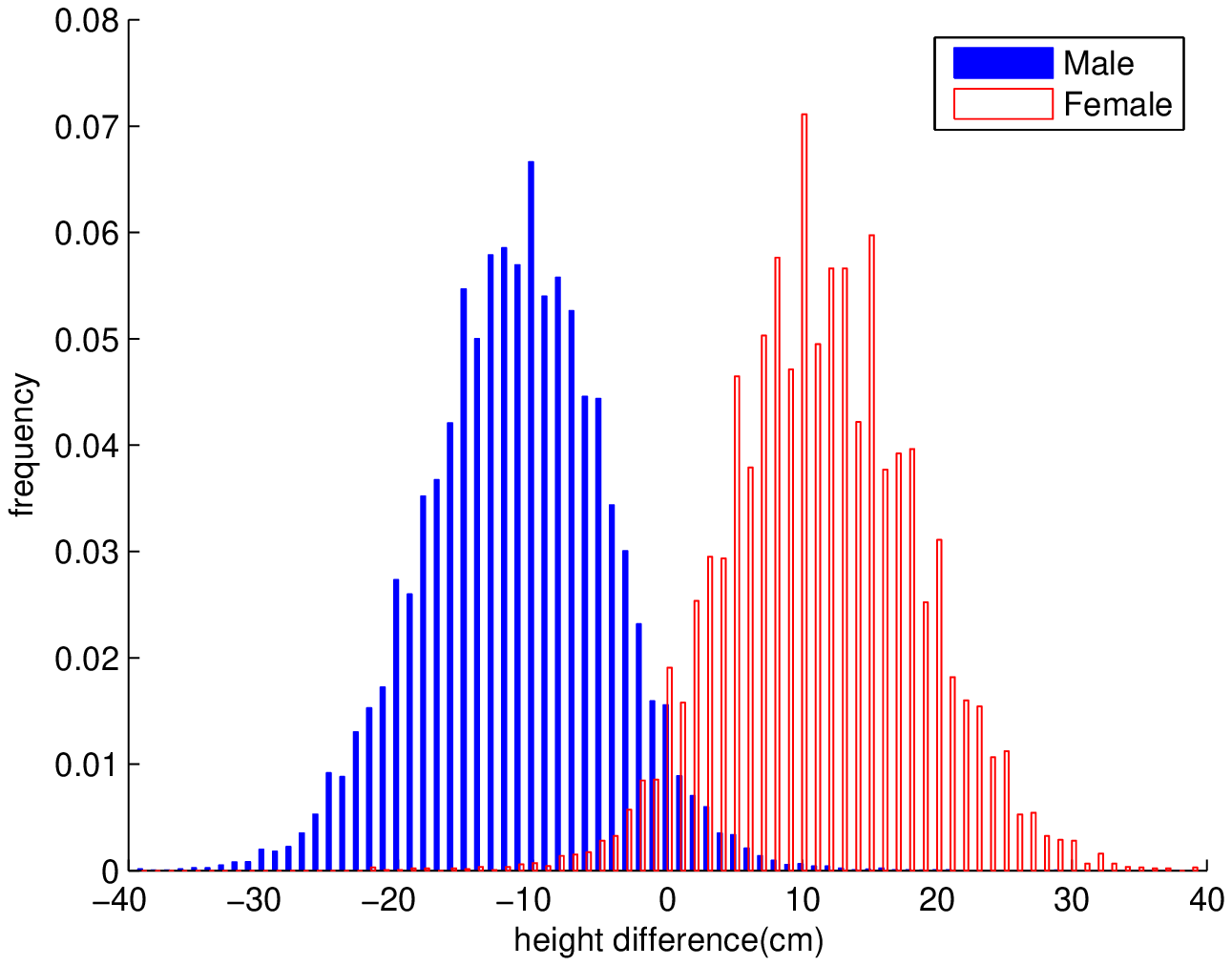}
		\label{fig:propertyDifferenceHeight}
	}
	\subfigure[Age]{
		\includegraphics[scale=0.34]{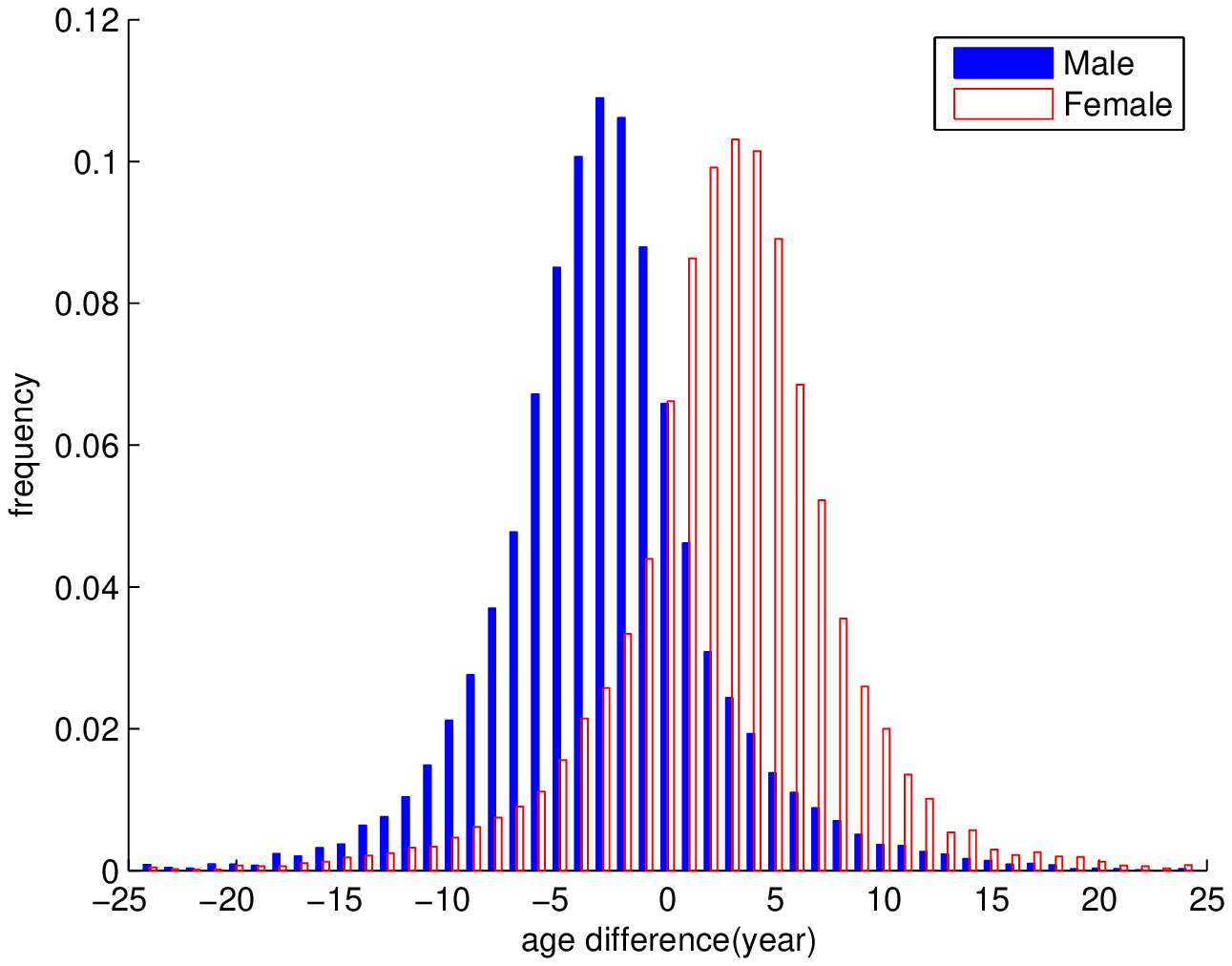}
		\label{fig:propertyDifferenceAge}
	} 
	\caption{
		Distribution of difference in height and age. 
		($\mathcal{M} = 29,274$, $\mathcal{F}= 14,981$.)
	}
	\label{fig:propertyDifference}
\end{center}
\end{figure}

The data enables us to quantify this.
We restrict the data to the ones that \emph{touch} to each other, 
that is,
one sends a virtual gift and the other responds to this gift.
Therefore both parties accepted the interaction.
This restriction reduces the data set to 
$\mathcal{F} = 14,981$ female and 
$\mathcal{M} = 29,274$ male. 

Consider an individual.
The individual touches many people.
The average height of all the people that 
the individual touches is calculated. 
The difference of the average and the height of the individual are obtained.
If the individual has no tendency for taller or shorter friend,
this difference is expected to be around zero.
If it is positive than the individual prefers friends that are taller.
In order to see if there is a difference due to gender,
individuals are grouped by gender.
Than the individual tendencies are averaged by each gender group.

\reffig{fig:propertyDifferenceHeight} provides tendency difference between genders.
The difference distribution is bell shape as expected.
The interesting observation is that the distributions are shifted with respect each other.
The mean of the distribution is around $-10$ cm for men,
whereas the mean for women is around $+10$ cm.
So the conclusion is that men prefers shorter women. 
On the other hand women prefers taller men.
It seems the ideal height difference between a man-women couple 
is expected to be around $10$ cm.

The distributions of men and women are similar in shape.
They seems to be symmetric around the mean.
Although both extends to $-20$ to $+20$ around the mean,
the frequency drops sharply as one goes away from the mean.
For example, 
men touching women taller than themselves are less than 5 \%.
Similarly, women preferring men shorter than themselves is also less than 5 \%.

\begin{figure}%[ht]
\begin{center}
	\subfigure[Education]{
		\includegraphics[scale=0.34]{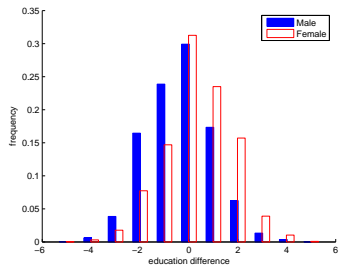}
		\label{fig:propertyDifferenceEducation}
	}
	\subfigure[Salary]{
		\includegraphics[scale=0.34]{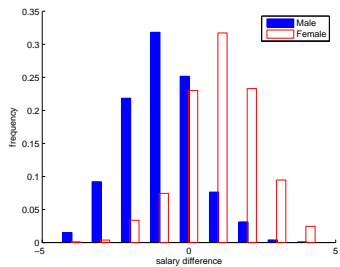}
		\label{fig:propertyDifferenceSalary}
	} \\
	\subfigure[BMI]{
		\includegraphics[scale=0.34]{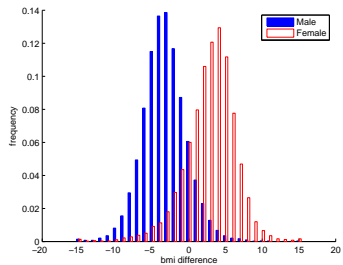}
		\label{fig:propertyDifferenceBMI}
	}
	\subfigure[Body Type]{
		\includegraphics[scale=0.34]{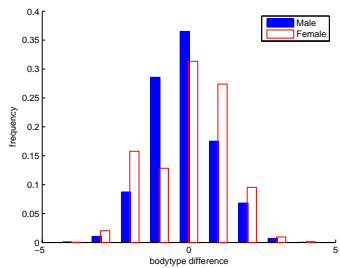}
		\label{fig:propertyDifferenceBodyType}
	} 
	\caption{
		Distribution of difference in educational level, salary, BMI and body type. 
		($\mathcal{M} = 29,274$, $\mathcal{F}= 14,981$.)
	}
	\label{fig:propertyDifference}
\end{center}
\end{figure}

Similar approach is taken for other properties.
\reffig{fig:propertyDifferenceAge},  
\reffig{fig:propertyDifferenceEducation}, 
\reffig{fig:propertyDifferenceSalary}, 
\reffig{fig:propertyDifferenceBMI}, and 
\reffig{fig:propertyDifferenceBodyType} 
provide 
gender differences in 
age,
education,
salary,
body mass index (BMI), and
body type, respectively.

\hCountDiff
In all the distributions, one observes a similar  shift due to gender. 
Men prefers negative, women prefers positive differences.

One should note that due to the design of the system,
properties have two different natures.
For some fields such as age and height, user is allowed to enter the exact value.
For example, someone of height $182$ cm, can enter exactly that value.
For some other fields, user selects a range.
Education, salary, body mass index (BMI), and body type are such fields.
For example for the salary field, 
user is allow to select one out of five salary bins, namely, 
bin 1: $x<500$, 
bin 2: $500<x<1000$,
bin 3: $1000<x<2000$,
bin 4: $2000<x<3000$, and
bin 5: $3000<x$.
So difference in the salary actually means 
the difference between the bin numbers not the actual salaries.
Fortunately, 
the mapping from the actual value to the bin number is order-preserving mapping.
Hence, the difference still has a meaning.

\hCountDiff
Note a very interesting point 
that both genders prefer the same amount of difference in all properties.
Consider a property such as height.
Suppose one gender prefers $10$ cm of height difference 
while the other $15$ cm.
Then, very few of us would be happy.
This is, perhaps,  due to an evolutionary adaptation 
which calls for an evolutionary explanation similar to that of ref~\cite{Silverman1992}.

\subsection{Distribution of Wink Messages}

\begin{table}%[htdp]
	\caption{Predefine wink messages.}
	\begin{center}
	\begin{tabular}{|c|p{6.5cm}|}
	%{|r|l|}
		\hline
		No	&Message\\
		\hline
		1	&I loved your photo. You are cool!\\
		2	&I liked your photo. You smile nice!\\
		3	&I liked your profile. Why don’t you have \\
			&a photo?\\
		4	&I liked your profile. Tell me about \\
			&yourself.\\
		5	&We are suitable for each other. Look at my \\
			&profile.\\
		6	&Everything starts with a hello. Hello!\\
		7	&If you’re looking for a funny friend, I am here.\\
		8	&I don’t want to add a message.\\
		\hline
	\end{tabular}
	\end{center}
	\label{tbl:winkList}
\end{table}%

The system provides a set of predefined messages, called \emph{wink}, 
that a user can send to.
The predefine wink messages are given in \reftbl{tbl:winkList}.
$\mathcal{M} = 216,891$ male and 
$\mathcal{F} = 241,748$ female users are included in this study.

\begin{figure}%[ht]
\begin{center}
	\subfigure[Education]{
		\includegraphics[scale=0.34]{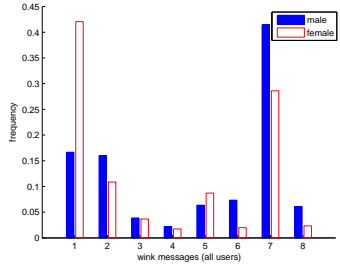}
		\label{fig:imgWinkMessageAll}
	}
	\subfigure[Salary]{
		\includegraphics[scale=0.34]{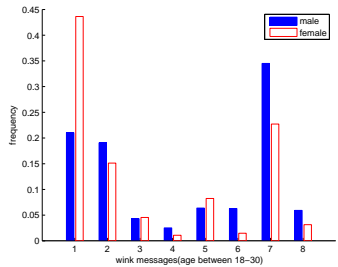}
		\label{fig:imgWinkMessage18-30}
	} \\
	\subfigure[BMI]{
		\includegraphics[scale=0.34]{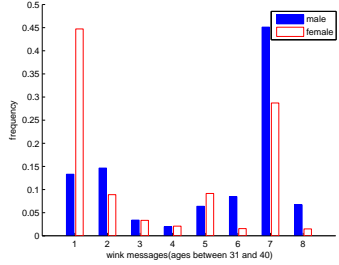}
		\label{fig:imgWinkMessage31-40}
	}
	\subfigure[Body Type]{
		\includegraphics[scale=0.34]{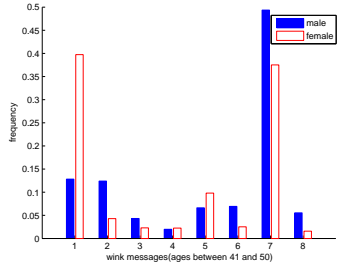}
		\label{fig:imgWinkMessage41-50}
	} 
	\caption{
		Distribution of wink message usage. 
		($\mathcal{M} = 216,891$, $\mathcal{F} = 241,748$.)
	}
	\label{fig:winkMessage}
\end{center}
\end{figure}

\hCountDiff
Usage of wink itself has asymmetry.
Male users use winking 10 times more than female users.
This can be interpreted as male takes the first move.

Selection of the wink message is also asymmetric as seen in
\reffig{fig:winkMessage}.
\reffig{fig:imgWinkMessageAll} provides message selection for all ages combined.
The message 
\emph{``7: If you’re looking for a funny friend, I am here.''}
is the most preferred message by male users. 
Interestingly, in almost one of every two winks by men, 
this message is selected. 
For female users this message is the second most preferred message.
The most frequent message of female users is 
\emph{``1: I loved your photo. You are cool!''}
with more than $40 \%$ usage.

It is observed that selection changes by age in both gender.
\reffig{fig:imgWinkMessage18-30}, 
\reffig{fig:imgWinkMessage31-40}, and 
\reffig{fig:imgWinkMessage41-50} provide change in selection 
for different age groups of $18-30$, $31-40$, and $41-50$, respectively.
As men gets older, they tent to use message 
\emph{``7: If you’re looking for a funny friend, I am here.''} more.
Its frequency goes up from $35 \%$ to almost $50 \%$.
There is no clear second message.
The second and the third frequently used messages,
namely,
\emph{``1: I loved your photo. You are cool!''} and
\emph{``2: I liked your photo. You smile nice!''}
are used almost at the same frequencies.

For the female users,
the most frequently used message is 
\emph{``1: I loved your photo. You are cool!''}.
As women get older, its frequency drops nearly from $45 \%$ to $40 \%$.
In contrast to males, 
there is a clear second most frequently used message for female users. 
It is
\emph{``7: If you’re looking for a funny friend, I am here.''}, 
which was the favorite of males.
Its usage increases approximately from $23 \%$ to $37 \%$ as women gets older.

For the mature age group of $41-50$, 
the first two messages hold more than $75 \%$ usage for females 
where as the same number is much lower, around $63 \%$ for males.
At all age groups, 
the coverage of the first three messages is much higher for females than males, 
that is, 
above $80 \%$ and $75 \%$, respectively.

\hCountDiff
This can be summarized as, 
visual appearance is more important than profile.
``Fun'' is always important and 
becomes increasingly more important for both genders as they get older.
Women decisively prefers to have ``cool photo'' as a initial message for friendship attempt 
as oppose to men prefers equally ``cool photo'' and ``nice smile'' messages.

\section{Discussion}

In many respects men and women behave differently~\cite{CelaConde2009PNAS}.
In the virtual world of online dating is another manifestation of this difference.
It is found that men take the first move in partner selection. 
Then, women have the right to accept or reject. 
Men are more open to share their private information then women. 
While male prefers women with lower qualifications such as income level, 
women do the opposite. 
Women prefer men of higher qualification such as higher education level. 
Both genders like funny partner yet the level of impact differs by gender and 
also by age with in the same gender.

\textbf{Acknowledgments.}
This work was partially supported 
by Bogazici University Research Fund, BAP-2008-08A105, 
by the Turkish State Planning Organization (DPT) TAM Project, 2007K120610,
and
by COST action MP0801.

\bibliographystyle{unsrt}
\bibliography{BingolGenderDifference}{}

% ====
\end{document}